\documentclass[journal]{IEEEtran} 

\normalsize

\ifCLASSINFOpdf
\else
\fi
\usepackage{setspace} 

\usepackage{cite}
\usepackage{ragged2e}
\usepackage{amssymb}
\usepackage{paralist}
\usepackage{amsmath}
\usepackage{amsthm}
\usepackage{graphicx}
\usepackage{epstopdf}
\usepackage{color}
\usepackage{extarrows}
\usepackage{multicol}
\usepackage{caption}
\usepackage{url}
\usepackage{bm}
\usepackage{footnote}
\usepackage{diagbox}
\usepackage{multirow}
\usepackage{array}
\usepackage{diagbox}
\usepackage{cases}

\usepackage[linesnumbered,ruled,vlined]{algorithm2e}
\usepackage{algorithmicx}

\usepackage[justification=centering]{subcaption}
\usepackage{booktabs}

\allowdisplaybreaks
\PassOptionsToPackage{bookmarks=false}{hyperref}

\def\be{ \begin{eqnarray} }
\def\ee{ \end{eqnarray} }
\usepackage{soul}
\usepackage{stfloats}

\begin{document}

\title{Digital Twin Assisted Intelligent Network Management for Vehicular Applications}

\author{Kaige Qu, \IEEEmembership{Member, IEEE}, and Weihua Zhuang, \IEEEmembership{Fellow, IEEE}
\thanks{This paper has been accepted by IEEE Wireless Communications.}
\thanks{This work was financially supported by research grants from the Natural Sciences and Engineering Research Council (NSERC) of Canada. \emph{Corresponding author: Kaige Qu.}}
\thanks{Kaige Qu and Weihua Zhuang are with the Department of Electrical and Computer Engineering, University of Waterloo, Waterloo, ON, Canada, N2L 3G1 (emails: \{k2qu, wzhuang\}@uwaterloo.ca).}
}

\maketitle

\begin{abstract}

The emerging data-driven methods based on artificial intelligence (AI) have paved the way for intelligent, flexible, and adaptive network management in vehicular applications. To enhance network management towards network automation, this article presents a digital twin (DT) assisted two-tier learning framework, which facilitates the automated life-cycle management of machine learning based intelligent network management functions (INMFs). Specifically, at a high tier, meta learning is employed to capture different levels of general features for the INMFs under nonstationary network conditions. At a low tier, individual learning models are customized for local networks based on fast model adaptation. Hierarchical DTs are deployed at the edge and cloud servers to assist the two-tier learning process, through closed-loop interactions with the physical network domain. Finally, a case study demonstrates the fast and accurate model adaptation ability of meta learning in comparison with benchmark schemes.  

\end{abstract}


\IEEEpeerreviewmaketitle

\section{Introduction}
\label{sec:Introduction}

\begin{figure}
\centering
{
\includegraphics[width=1\linewidth]{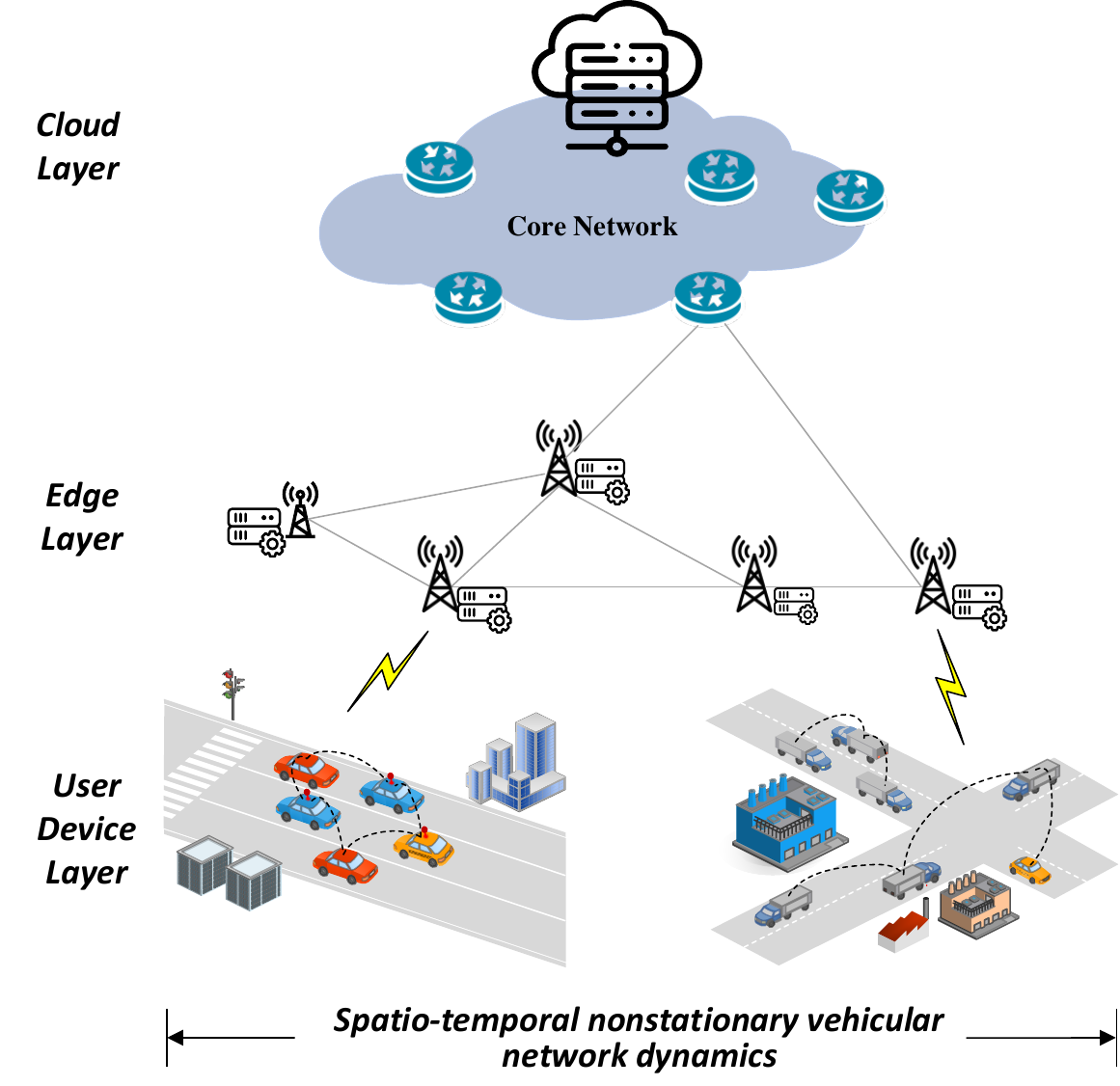}} 
\caption{A three-layer network architecture for vehicular networks.}\label{fig:Two-Tier-Learning_1}
\end{figure}

The sixth-generation (6G) wireless communication networks are foreseen to support emerging vehicular applications that will dramatically boost vehicle intelligence and revolutionize the transportation systems, such as autonomous driving and augmented reality (AR) entertainment for onboard passengers~\cite{10250875}. 
These applications usually rely on powerful computing servers to meet real-time performance requirements. 
A layered network architecture spanning across the user end device, edge, and cloud layers, as illustrated in Fig.~\ref{fig:Two-Tier-Learning_1}, supports such vehicular applications by exploiting the edge and cloud computing resources. 
In the network architecture, there should be various network management functions for different network optimization purposes, such as to coordinate the multi-dimensional resources for computing, communication, sensing, positioning, and storage, to control the task offloading process from vehicle users to edge/cloud servers for delay satisfaction, and to support the mobility of vehicle users via handover and service migration~\cite{wu2022ai,shen2021holistic}. 
Formally, network management involves the monitoring, configuration, analysis, evaluation and control of the network elements and resources to meet service quality requirements at a reasonable cost~\cite{coronado2022zero}.

High dynamics in a vehicular network, such as vehicle density, data traffic volume, computing demand, sensing workload and channel conditions, complicate the network management. 
To handle such complexity, cost-effective scalable performance optimization solutions and adaptive algorithms are required for intelligent and automated network management~\cite{benzaid2020ai}. 
The traditional model-driven approaches follow the “model-then-optimize” manner and rely on domain expert knowledge. They have the advantage of generalities for different networking scenarios, but are limited to simple or small-scale networks, thus becoming inadequate and even intractable in 6G. 
Emerging data-driven approaches rely on artificial intelligence (AI) and big data, bringing a promising alternative for solving complex network optimization problems.
To enhance the overall intelligence of vehicular networks, AI techniques can be employed to embed intelligence in network management, creating \emph{intelligent network management functions} (INMFs)~\cite{coronado2022zero}. 

The INMFs can potentially capture system dynamics and facilitate the intelligent network management. 
However, innovative engineering solutions are required towards network management automation. 
Typically, a machine learning (ML) model converges once it has successfully captured a stationary distribution in the training data, and a trained model can be used for real-time network management under the assumption that the underlying data distribution does not change. 
However, the assumption is not true if 
we consider a vehicular network over a long time period or in a large geographical area, where network dynamics exhibit spatio-temporal statistical variations~\cite{yuan2022ai}.  
For example, for different road conditions and time periods, the vehicle density distributions are different, and the resource demand can have significant changes. In such cases, the trained ML models can perform poorly, 
and a model retraining should be triggered to customize the INMFs for the new scenarios~\cite{3gpp.23.700-81}. Typically, retraining from scratch is time-consuming and data-inefficient~\cite{benzaid2020ai}. 
A question is how to achieve fast and efficient model adaptation to overcome the spatio-temporal nonstationarity in a vehicular network.  

\begin{figure}
\centering
{
\includegraphics[width=0.95\linewidth]{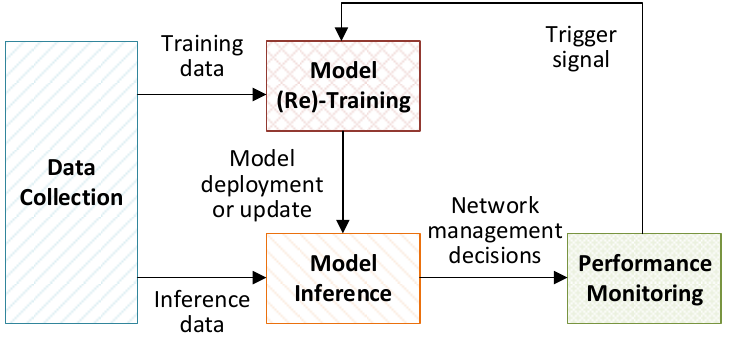}} 
\caption{Stages in the life-cycle of a machine learning model.}\label{fig:Two-Tier-Learning_5}
\end{figure}

In this study, we take a step further from data-driven network management towards network automation. 
The former involves \emph{individual learning tasks} for developing customized ML models of different INMFs in a stationary network scenario.  
The latter further involves the automated life-cycle management of such ML models under nonstationary conditions. As illustrated in Fig.~\ref{fig:Two-Tier-Learning_5}, the life-cycle of a machine learning model includes data collection, model training and re-training, model inference, and performance monitoring. The different phases should be automatically triggered and managed, to enable learning task evolution in a nonstationary network environment~\cite{3gpp.23.700-81,benzaid2020ai,ajayi2023self}. 
Meta learning, also known as ``learning to learn'', aims at learning how to efficiently execute new learning tasks. 
The key idea is to leverage prior learning experiences to improve the learning process itself and to create \emph{meta models} that can quickly adapt to new learning tasks with limited data~\cite{finn2017model,wang2022meta}. 
Here, we discuss meta learning based network management for the spatio-temporal nonstationary vehicular networks, and propose a two-tier learning framework in the three-layer network architecture. 
Such a framework aims at providing flexibility in learning model life-cycle management, as an initial step towards network automation.

Recently, digital twin (DT) has captured significant attention due to its potential to transform industries by providing a bridge between the physical and digital worlds  and including its role in vehicular networks~\cite{shen2021holistic,cai2023task,guo2023five}. 
A DT refers to a virtual representation of a physical system. 
It should be continuously updated with real-world data to reflect the current state and behavior of the physical counterpart, which allows for simulation, analysis, monitoring, and optimization of the physical system without actually interacting with it.
Thus, DTs can potentially assist in our two-tier learning framework. 
We consider hierarchical DTs, including lower-level DTs deployed at edge servers for individual model learning and higher-level DTs deployed at cloud servers for meta learning.  

The remainder of this article is organized as follows. We first introduce preliminaries of fast model adaptation based on meta learning, and then propose the DT-assisted two-tier learning framework. A case study is presented to demonstrate the benefit of meta learning, followed by a conclusion.

\section{Preliminaries of Fast Model Adaptation} 
\label{sec:Preliminaries of Fast Model Adaptation}

For learning task evolution under a nonstationary network condition, when to trigger model retraining and how to achieve fast model adaptation are critical issues. 
In theory, a model retraining is required if the system dynamics experience a distribution drift. However, as it is difficult to accurately estimate an unknown probability distribution, we can monitor the model performance and use the performance degradation as a triggering signal in practice. 

An ideal model retraining should be fast and accurate to adapt to new network conditions. 
Retraining from scratch is not a good choice, as the model convergence is slow especially when new training data are scarce. A long retraining time deteriorates the robustness of ML-based network optimization solutions. 
The old model is not a good initial point either, as it might be overfitted to the old data, which possibly slows down the model retraining especially if the new data distribution significantly deviates from the old one.  
A desired initial point is a model with good generalization ability. 
Such a model is referred to as a \emph{meta model} in the context of meta learning. 
The core concepts of meta learning include: 
\begin{itemize}

	\item \emph{Meta training}: A meta model is trained by learning from a diverse set of individual learning tasks, to learn a high-level initialization that quickly adapts to new tasks; 

	\item \emph{Meta adaptation}: The trained meta model serves as an initial point for fast model adaptation on new learning tasks with a limited amount of data.
	Such a few-shot learning capability makes meta learning particularly useful in scenarios where data are limited. 

\end{itemize}

Meta learning can be applied in learning frameworks such as supervised learning and reinforcement learning (RL)~\cite{finn2017model,wang2022meta}. 
We briefly introduce the meta training procedures in the context of RL. 
Consider an unknown probability distribution over Markov decision process (MDP) tasks for a specific INMF such as resource allocation, task offloading, mobility management, and packet scheduling in vehicular networks. 
An individual learning task corresponds to a random MDP sampled from such a task distribution, which has unknown but stationary state transition probabilities. 
RL can be used to solve the MDP, and the RL model is referred to as an \emph{individual model}. 
A meta model has the same neural network structure as the individual models, but with a set of different parameters. 
The meta model aims to learn general features broadly applicable to all MDP tasks sampled from the task distribution, thus can adapt well among different tasks.

Many meta learning algorithms, such as model-agnostic meta learning, use gradient-based optimization techniques that iteratively update meta model parameters to minimize an average adaptation loss on sampled tasks~\cite{finn2017model}.  
For each sampled task, the adaptation loss captures the ability of the meta model to quickly adapt its learned knowledge and perform well on the task. 
In each iteration, a number, $I$, of individual learning tasks are sampled, and the following steps are performed between one meta learning agent and $I$ individual learning agents. 
\begin{itemize}

	\item \textbf{Individual model initialization}: The meta learning agent distributes the current meta model to the $I$ individual learning agents for individual model initialization; 

	\item \textbf{First-round data collection}: Each individual learning agent collects a set of trajectory data composed of state-action-reward transitions by using the current individual model. They either interact with the real vehicular network environment to obtain the true data or emulate the network operations to obtain the synthetic data; 

	\item \textbf{Individual model adaptation}: With the trajectory data, each individual learning agent performs an individual model adaptation via one gradient descent over the loss function of the individual model (i.e., individual loss); 

	\item \textbf{Second-round data collection}\footnote{In some meta learning algorithms such as Reptile, this step is omitted by approximating the adaptation loss gradients as the scaled difference between meta model parameters and adapted individual model parameters~\cite{nichol2018first}.}: In order to evaluate the adaptation loss of each sampled task, each individual learning agent collects an extra set of trajectory data by using the adapted individual model;

	\item \textbf{Adaptation loss evaluation}: For each individual learning agent, the adaptation loss is the individual loss with the new trajectory data, and the adaptation loss gradients are computed with regards to the meta model parameters and sent to the meta learning agent; 

	\item \textbf{Meta model update}: Once all $I$ sets of adaptation loss gradients are available, the meta learning agent updates the meta model via a gradient descent in the direction of the average adaptation loss gradients, to minimize the average adaptation loss on the $I$ sampled tasks.

\end{itemize}

A meta model is trained once the average adaptation loss on sampled tasks converges after multiple iterations. 
Then, for any individual learning task from the same task distribution, a customized individual model can be trained via meta adaptation, which may take several gradient descent steps. 

\section{A Digital Twin Assisted Two-Tier Learning Framework}
\label{sec:A Digital Twin Assisted Two-Tier Learning Framework}

\begin{figure*}
\centering
{
\includegraphics[width=1\linewidth]{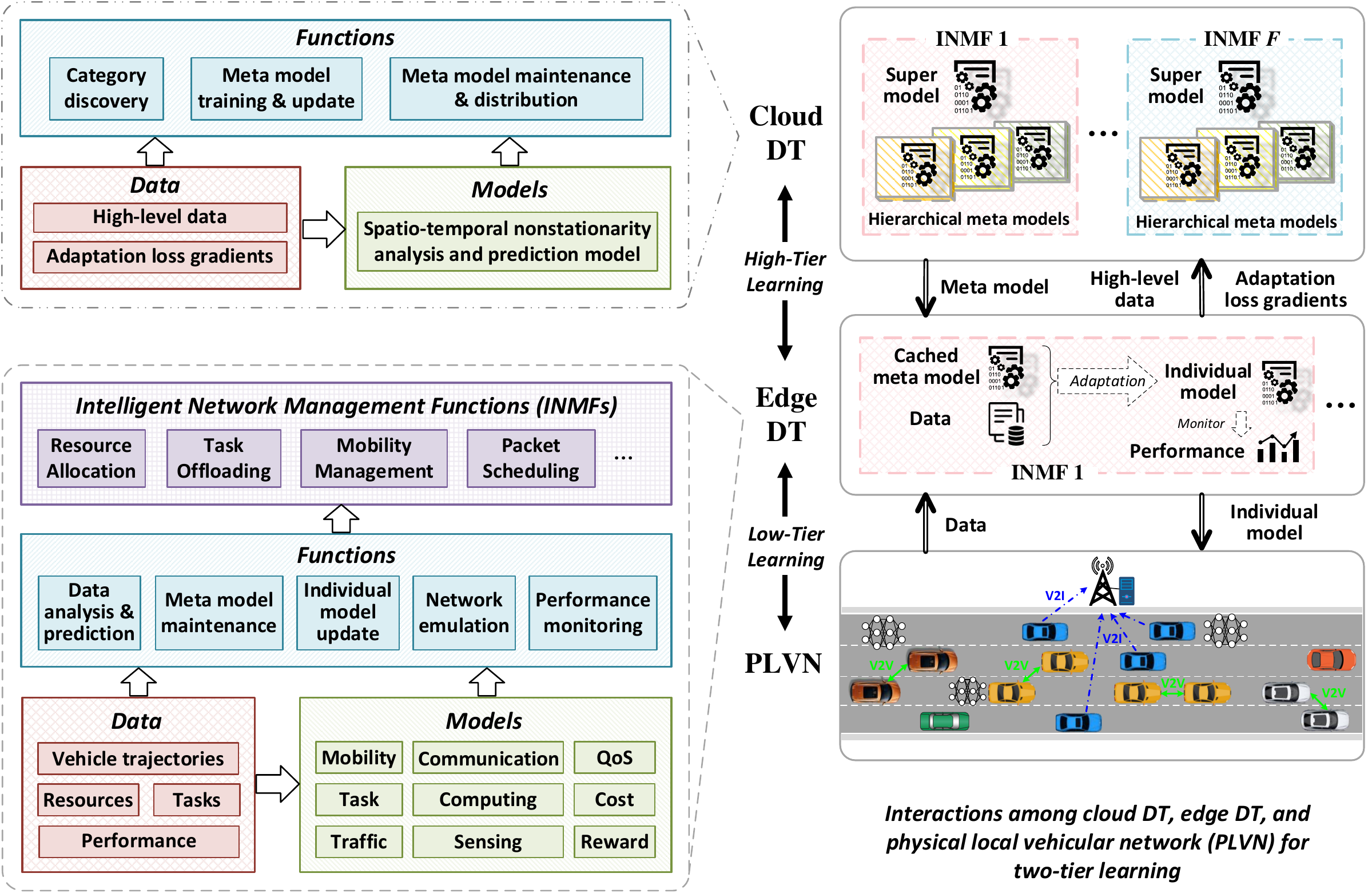}} 
\caption{A digital-twin assisted two-tier learning framework for vehicular networks.}\label{fig:Two-Tier-Learning_17}
\end{figure*}

We consider a number of \emph{physical local vehicular networks} (PLVNs) with nonstationary spatio-temporal characteristics in a geographical region. 
For each PLVN, a set of ML models should be customized to support multiple INMFs. 
For each INMF, a PLVN is associated with one individual model, and meta learning is used to capture the general features among the PLVNs. 
For such meta learning based network management, we propose a digital twin (DT) assisted two-tier learning framework in the three-layer network architecture, as illustrated in Fig.~\ref{fig:Two-Tier-Learning_17}.   
\emph{First}, we consider hierarchical meta models with different generalization levels. 
\emph{Second}, we propose hierarchical DTs to support two-tier learning. 
\emph{Third}, we discuss the offline planning and online operation stages of the proposed framework, which enable closed-loop interactions among the cloud, edge, and end device layers, and between the hierarchical DTs and the physical network domain.

\subsection{Hierarchical Meta Models}

\begin{figure}
\centering
{
\includegraphics[width=1\linewidth]{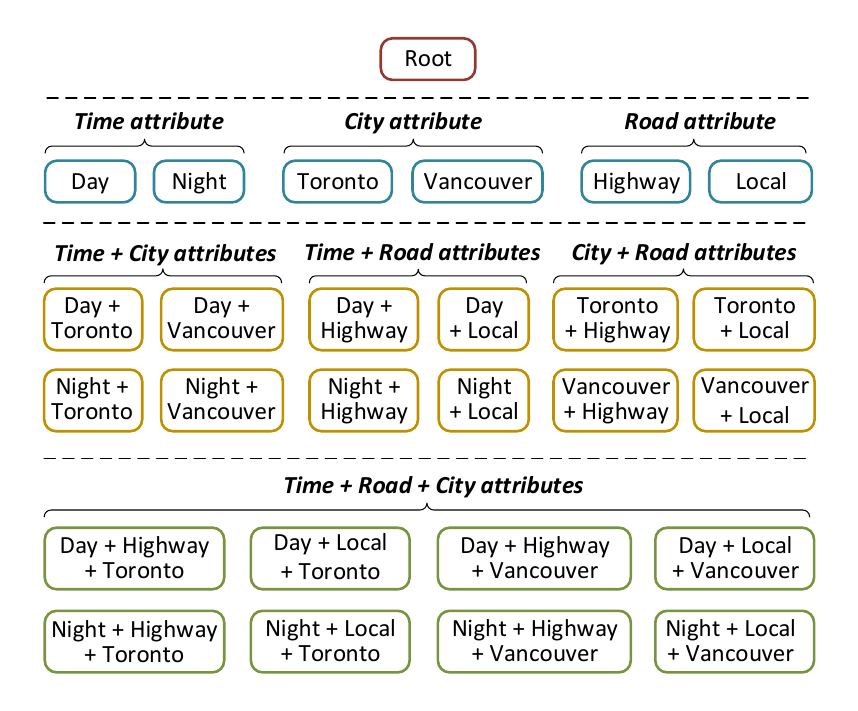}} 
\caption{An illustrative example of hierarchical meta models for different network categories based on time, city, road attributes.}\label{fig:Two-Tier-Learning_18}
\end{figure}

For each INMF, a single meta model trained based on random PLVNs is to capture the most high-level features shared among the networks. The ``distance'' from such a meta model to any individual model is not close due to a very high generalization level, leading to a long model adaptation delay. 
To balance between generalization and customization, we propose \emph{hierarchical meta models} for each INMF to capture different levels of general features among the PLVNs. 
Specifically, we extract network-level attributes (such as hour, city, location, road condition, vehicle number, average workload, and average resources) of the PLVNs, all of which have an impact on the unknown network dynamics models. 
Based on one attribute or a combination of attributes, the PLVNs fall into different categories. For example, a network can be tagged by a time attribute such as ``Day'' and ``Night'', or a combination of time, city and road attributes such as ``Day + Toronto + Highway'' and ``Night + Vancouver + Local''. 
To maintain a list of PLVNs for each category, a unique ID is assigned to each PLVN. 
A meta model is trained for each category to learn the general features within the corresponding PLVN list. 

The attributes should be carefully selected and processed. 
More attributes lead to more fine-grained categories and less-general meta models with a higher customization level. Such meta models require less fine-tuning, enabling faster model adaptation. 
However, the data inefficiency issue in training the meta models arises, as it takes a longer time to gather sufficient training data from each fine-grained category. 
Additionally, for the categorization, discretization is needed for continuous attributes. 
We can normalize those with different units/scales to the range of $[0,1]$ before discretization.

Fig.~\ref{fig:Two-Tier-Learning_18} illustrates hierarchical meta models for different categories based on time, city and road attributes. 
The root node represents a super (meta) model that captures the most general features among all PLVNs.

\subsection{Hierarchical Digital Twins}

As shown in Fig.~\ref{fig:Two-Tier-Learning_17}, hierarchical DTs are deployed at the edge and cloud layers to assist two-tier learning based network management. 
A cloud DT is created for high-tier learning among all PLVNs. 
It trains and maintains sets of hierarchical meta models for multiple INMFs. 
For each PLVN, an edge DT is created for low-tier learning, specifically for individual model adaptation and customization based on meta models from the cloud DT. 
The low-tier learning is necessary for both meta training and meta adaptation, as discussed in Section~\ref{sec:Preliminaries of Fast Model Adaptation}.  

An \emph{edge DT} is a digital copy of a PLVN, which can be deployed at the nearest edge server. 
The edge DT has the same ID as the associated PLVN.   
If a PLVN corresponds to a moving cluster of vehicles, the associated edge DT migrates among edge servers to ``\emph{follow}'' the vehicle cluster, thus providing the mobility support. 
The data of the PLVN such as vehicle trajectories, task information, resource availability, QoS/cost measurements are periodically updated to the edge DT. The edge DT also maintains models of the PLVN for network emulation, which facilitates the coordination of data-driven and model-based network management~\cite{shen2021holistic}. Examples of the models include mobility model, task model, data traffic model, communication/computing/sensing models, and QoS/cost/reward models. Such models can be based on domain expert knowledge and enhanced by experience data. 

Based on the data and models, the edge DT can support multiple functions to facilitate the low-tier learning for diverse INMFs. For example, the network-level attributes of the associated PLVN can be analyzed based on the collected data.
With the attributes, the potentially changing PLVN categories can be determined over time. 
The ID and categories of each edge DT are recorded by the cloud DT. Then, an edge DT can be involved in a meta training process initiated by the cloud DT for meta models of matching categories. 
For each INMF requested by the PLVN, if there are multiple available meta models of matching categories at the cloud DT, the least-general one based on known attributes is fetched and then cached at the edge DT. 
With the cached meta model as an initial point, an individual model can be trained or updated to fit the current network conditions. 
Otherwise, the super model is used by default for individual model learning.

The \emph{cloud DT} is a high-level virtual representation of all the PLVNs.  
To support training a set of hierarchical meta models per INMF, the cloud DT collects high-level data (such as network-level attributes) and adaptation loss gradients from the edge DTs. 
It also maintains models that capture the overall system characteristics. 
For example, based on the high-level data, the spatio-temporal nonstationarity among the PLVNs can be analyzed. For a PLVN, the change points in time between stationary network conditions can be detected, and the new network conditions can be predicted~\cite{9171912}.   
Moreover, a drift of the unknown learning task distribution within each PLVN category can be coarsely detected. 
For each category, the fractions of PLVNs in all the sub-categories composite a vector. The Euclidean distance between vectors at different time points roughly evaluates the difference in the task distribution. 
Additionally, a remarkable increase of the meta model's adaptation loss on sampled tasks implies a task distribution drift.   
Based on such high-level analyses, meta model training for new categories and meta model update for existing categories can be triggered.

\subsection{Offline Planning and Online Operation Stages}

\begin{figure*}
\centering
{
\includegraphics[width=0.95\linewidth]{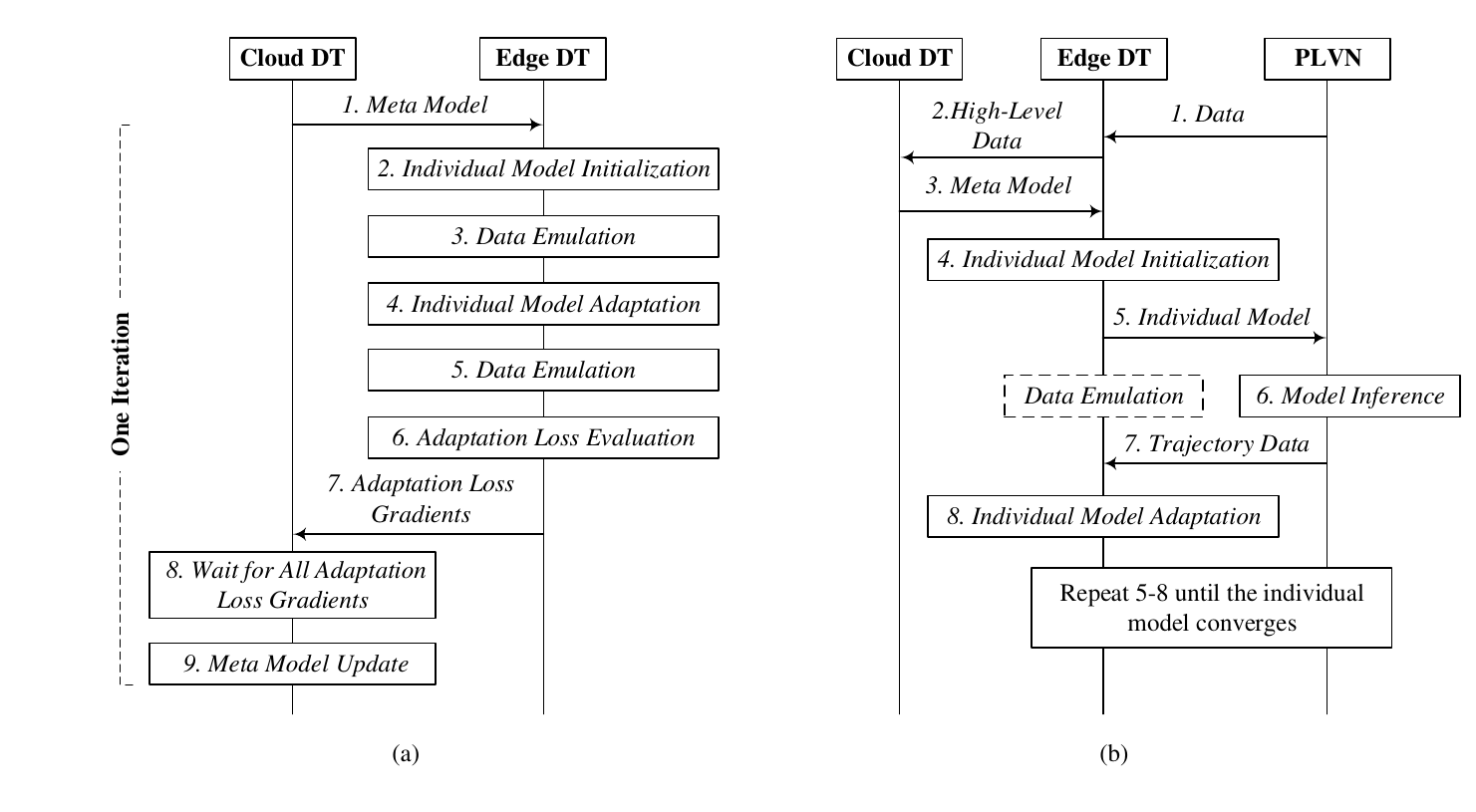}} 
\caption{Interactions among cloud DT, edge DT, and PLVN. (a) for meta training. (b) for meta adaptation. }\label{fig:Interactions}
\end{figure*}

\subsubsection{Offline planning stage}

This stage can take a long time duration, e.g., several days or weeks. 
Based on a pre-determined set of attributes, network categories with different granularities can be determined. 
For each INMF, a meta model is trained for each category at the cloud DT in multiple iterations. 
In each iteration, a set of $I$ PLVNs are uniformly sampled from the category's PLVN list, and the current meta model is distributed to the associated edge DTs. 
With the maintained PLVN list, the PLVN sampling can be performed without explicitly knowing the learning task distribution. 
The edge DTs evaluate the meta model with emulated data trajectories, and compute the adaptation loss gradients which are then returned to the cloud DT for meta model update. 
Fig.~\ref{fig:Interactions}(a) illustrates the interactions between the cloud DT and one edge DT for meta training in one iteration. 
There are two data emulation steps before and after the individual model adaptation. 
Conventionally, to train an individual model, data trajectories are acquired from the physical network. 
However, as the individual model adaptation here is for adaptation loss evaluation rather than for formally training an individual model, the network evolution can be emulated to generate synthetic trajectory data based on the data and models in the edge DT, which helps to avoid poor network performance during the long meta training process. 
Once trained, the hierarchical meta models for each INMF are stored in the cloud DT. 
To save the computation and storage resources for training and maintaining the meta models, techniques such as multi-task learning can be employed to enable representation sharing among highly-related INMFs. 

We do not store individual models in the cloud DT, although a transfer learning technique allows model adaptation from a source individual learning task to a related target task. The reason is that it is difficult to accurately match the PLVNs to existing individual models due to the unavailability of accurate network dynamics models. 
Correlation analysis based on high-level attributes may help to evaluate the similarity between PLVNs, but its validity and accuracy need more investigation. 
Such analysis also incurs computation overhead. 
In comparison, it is more convenient and cost-effective to identify the categories of any PLVN, and find a proper meta model for fast individual model customization.

\subsubsection{Online operation stage}

During this stage, a customized individual model is created for each INMF of a PLVN. 
Based on the data collected from the PLVN, the edge DT extracts and sends high-level data such as attributes to the cloud DT. 
If the PLVN belongs to a known category, a meta model is fetched from the cloud DT for each INMF and then cached by the associated edge DT. 
The edge DT then customizes an individual model for each INMF through several meta adaptation steps, as shown in Fig.~\ref{fig:Interactions}(b). 
Each adaptation step requires an interaction between the edge DT and the PLVN. 
To accelerate the model convergence, a hybrid data collection mode can be employed, where true trajectory data are collected via model inference in the PLVN and synthetic trajectory data are generated in the edge DT.   
Once the individual models are trained, the edge DT dispatches them to the PLVN for real-time network management based on model inference. 
During the inference phase, the true trajectory data and performance measurements can be updated to the edge DT, to tune the network emulation models and trigger necessary individual model retraining. 

For a PLVN requiring model retraining, if it remains in the same category, a new individual model is retrained based on the cached meta model. 
If it transfers to another known category, a new meta model should be fetched from the cloud DT. 
If it transfers to an unknown category, the super model is used for individual model retraining, and a meta model for the new category is trained in the cloud DT. 
As the individual models are difficult to be matched to a PLVN, the old individual models are deleted once the new ones are trained, to save the caching space. 
Nevertheless, a list of meta models can be cached at the edge DT, as meta model fetching from the cloud DT incurs long propagation delay through the Internet.
For delay improvement, a cache replacement policy can be designed to keep the most popular meta models. 

The existing meta models (including the super model) maintained by the cloud DT should be updated if the respective task distributions experience significant changes, which can be fulfilled by the same meta training procedures as shown in Fig.~\ref{fig:Interactions}(a). 
Therefore, with the continuous online operation, a complete library of up-to-date meta models for different network categories and INMFs will be gradually established in the cloud DT, and the individual models are automatically created, updated, and deleted at the edge DTs for the spatio-temporal nonstationary PLVNs.

\section{Case Study}

We discuss one use case for RL-based adaptive cooperative perception among connected and autonomous vehicles (CAVs), and focus on the meta learning aspect. 
We consider a moving vehicle cluster in the service coverage of a road-side unit (RSU). The vehicle cluster consists of multiple CAV pairs which may perform cooperative perception via vehicle-to-vehicle (V2V) communication, along with human-driven vehicles which have potential vehicle-to-RSU transmission. 
Due to the radio resource sharing among all the vehicles, the radio resource availability for supporting the cooperative perception of CAV pairs is dynamic. Moreover, due to the vehicle mobility, the perception workloads (i.e., number of nearby objects for detection and classification) and the channel conditions for the CAV pairs vary with time. 
Under such network dynamics, each CAV pair can switch between a default \emph{stand-alone perception (SP) mode} and a selective \emph{cooperative perception (CP) mode}, to ensure consistent delay satisfaction~\cite{Qu-J7}. 
CP potentially reduces the total computing demand at a feature data transmission cost between CAVs. For delay satisfaction, 
the CPU frequency for supporting the computation in perception tasks should be scaled up/down on demand. 
Define the computing efficiency gain of a CAV pair as the reduced amount of computing energy consumption in comparison with that in the default SP mode. 
Such a gain is equal to zero in the SP mode, and decreases proportionally with computing demand (in CPU cycle) and CPU frequency (in cycle/s) squared in the CP mode~\cite{Qu-J7}.  

An increase of the perception workload at a CAV pair in the CP mode leads to more reduction in the total computing demand in comparison with the SP mode, at the cost of higher CPU frequency and transmission rate requirements due to a reduced per-object delay budget. 
The transmitter-receiver distances of the CAV pairs also affect the feature data transmission delay.
Hence, the dynamic network state in terms of the available radio resources for V2V transmission, the perception workloads of all CAV pairs, and the transmitter-receiver distances of each CAV pair should be considered in the adaptive perception mode selection, to maximize the total computing efficiency gain while satisfying the delay requirement. 
We formulate the adaptive cooperative perception problem as an MDP characterized by state, action, and reward. 
The action is a binary decision vector indicating the selection between the SP and CP modes among all CAV pairs. 
The reward is the total computing efficiency gain. 

We propose an RL solution based on a proximal policy optimization (PPO) algorithm. 
In PPO, we train a Gaussian stochastic policy network which learns the mean and the standard deviation of a set of Gaussian random variables for all CAV pairs. For each CAV pair, a continuous action is sampled from the corresponding Gaussian distribution and then discretized to a binary cooperation decision. 
The policy entropy is the average differential entropy of all the Gaussian random variables, which measures the amount of uncertainty or randomness in the stochastic policy. 

We conduct simulations to evaluate how meta learning accelerates the model adaptation. 
The RL agent interacts with a dynamic network environment including $3$ CAV pairs and $10$ HDVs in consecutive episodes. Each episode contains $K=75$ time slots. 
At time slot $k$, the amount of available radio resources follows a Normal distribution $\mathcal{N}\sim(\mu(k),\sigma^2)$, where $\mu(k)$ is the mean and $\sigma$ is the time-invariant standard deviation. 
The mean has three possible values in $\{5, 6, 7\}$ MHz, corresponding to $low$, $medium$, and $high$ resources, and $\sigma$ is set to $0.3$ MHz. 
An individual RL task is associated with a customized radio resource pattern in each episode, with mean values denoted by sequence $\{\mu(0),\cdots,\mu(K-1)\}$.  
We consider two individual RL tasks with $low$ and $high$ resource availability respectively. Specifically, mean $\mu(k)$ for each time slot is always $5$ MHz ($7$ MHz) for task 1 (task 2). A PPO model is trained from scratch (\textbf{PPO-random}) for both tasks. 
To train a meta model, we create a set of random network environments associated with random radio resource patterns, by uniformly sampling $\mu(k)$ among $\{5, 6, 7\}$ MHz. 
For task 1, two extra PPO models are trained based on meta adaptation (\textbf{PPO-ML}) and transfer learning from task 2 (\textbf{PPO-TL}), respectively.

\begin{figure}
    \centering
    \begin{subfigure}[b]{0.4\textwidth}
      \includegraphics[width=1\linewidth]{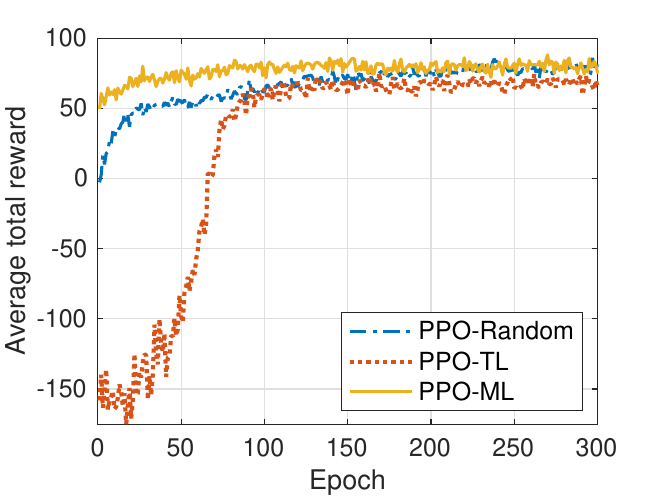} 
      \caption{}\label{fig:Comp_return_cvg}
    \end{subfigure}
    ~~~ 
    \begin{subfigure}[b]{0.4\textwidth}
  \includegraphics[width=1\linewidth]{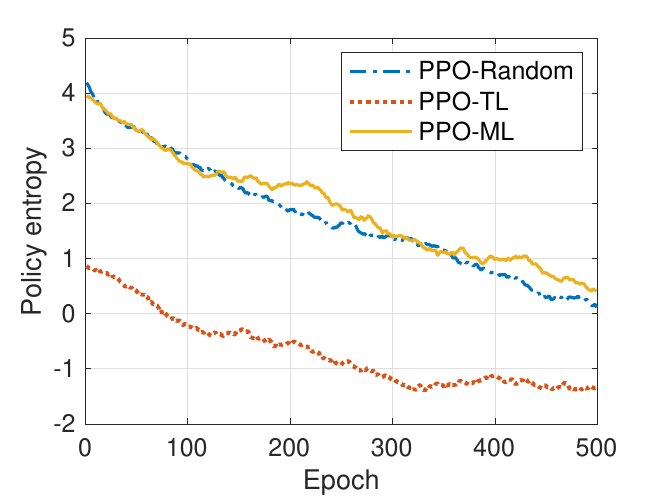} 
  \caption{}\label{fig:Comp_entropy}
    \end{subfigure}
  \caption{Performance comparison between training from scratch, transfer learning, and meta learning for task 1. (a) average total reward. (b) policy entropy. }\label{fig:comp}
\end{figure}

All the PPO models are trained over 500 training epochs, each consisting of $10$ episodes. 
Fig.~\ref{fig:comp} shows the per-episode average total reward and policy entropy over training epochs for the three PPO models of task 1. 
As shown in Fig.~\ref{fig:comp}(a), the meta model adapts quickly and well, as indicated by the fast convergence and the comparable average total reward with PPO-random after convergence.
In comparison, transfer learning shows inferior performance than meta learning, in terms of both slower convergence and lower average total reward after convergence. 
The potential reason for the superiority of meta learning can be inferred from Fig.~\ref{fig:comp}(b). 
As a meta model captures the general features among random learning tasks, its policy entropy is comparable to that of a random model, which encourages better exploration in the right direction during the model adaptation. 
However, in PPO-TL, as a trained model is customized to a task, the policy entropy is low due to less randomness. This hinders the exploration during the transfer learning process 
and makes the model more easily stuck at a local optimum, especially when the source and target tasks significantly differ from each other.

\section{Conclusion}

In this article, we present a digital twin assisted two-tier learning framework, to support the automated life-cycle management of AI-based intelligent network management functions for vehicular applications. 
The cloud digital twin builds a dynamic collection of hierarchical meta models, while the edge digital twin facilitates fast individual model customization. 
Such a framework aims at a better trade-off between generalization and customization for the intelligent network management in vehicular networks, providing a promising tool towards network automation. 
In our future work, we will further explore the strengths of digital twin and meta learning for more use cases in vehicular networks, and potentially extend our ideas to space-air-ground integrated networks. 

\section*{Acknowledgement}

The authors would like to thank the undergraduate research assistant, Chinemerem Chigbo, for implementing the meta learning algorithm.

\bibliographystyle{IEEEtran}
\bibliography{Ref_final}

\end{document}